\documentclass[pra,aps,tightenlines,twocolumn,superscriptaddress,showpacs,eqsecnum,floatfix]{revtex4}
\usepackage{epsfig,amsmath,amssymb,bm,epsf,graphics}
\def\mygamma{{r}}
\begin{document}
\title{Maximal entanglement versus entropy for mixed quantum states}
\author{Tzu-Chieh Wei}
\affiliation{Department of Physics, University of Illinois at Urbana-Champaign, 
1110 West Green Street, Urbana, Illinois 61801-3080, USA}
\author{Kae Nemoto}
\affiliation{Informatics, Bangor University, Bangor LL57 1UT, UK}
\author{Paul M.~Goldbart}
\affiliation{Department of Physics, University of Illinois at Urbana-Champaign, 
1110 West Green Street, Urbana, Illinois 61801-3080, USA}
\author{Paul G.~Kwiat}
\affiliation{Department of Physics, University of Illinois at Urbana-Champaign, 
1110 West Green Street, Urbana, Illinois 61801-3080, USA}
\author{William J. Munro}
\affiliation{Hewlett-Packard Laboratories, Filton Road, Stoke Gifford,
         Bristol, BS34 SQ2, UK}
\author{Frank Verstraete}
\affiliation{Department of Mathematical Physics and Astronomy,
Ghent University, Belgium}
  \date{August 21, 2002}
% \date{\today}
%
\begin{abstract}
Maximally entangled mixed states are those states that, for a given mixedness,
achieve the greatest possible entanglement. For two-qubit systems and for various combinations of entanglement and mixedness measures, the form of the corresponding maximally entangled mixed states is determined primarily analytically. As measures of entanglement, we consider entanglement of formation, relative entropy of entanglement, and negativity; as measures of mixedness, we consider linear and von Neumann entropies. We show that the forms of the maximally entangled mixed states can vary with the combination of (entanglement and mixedness) measures chosen. Moreover, for certain combinations, the forms of the maximally entangled mixed states can change discontinuously at a specific value of the entropy. 
\end{abstract}
\maketitle
\section{Introduction}
\label{SEC:introd}
Over the past decade, the physical characteristic of the entanglement  
of quantum-mechanical states, both pure and mixed, has been recognized as 
a central resource in various aspects of quantum information processing.  
Significant settings include quantum communication~\cite{Schumacher96},
cryptography~\cite{Ekert91},  
teleportation~\cite{Bennett_Brassard_Crepeau_Jozsa_Peres_Wootters}, and,
to an extent that is not quite so clear, quantum computation~\cite{DiVincenzo95}.
Given the central status of entanglement, the task of quantifying the 
degree to which a state is entangled is important for quantum information 
processing and, correspondingly, several measures of it have been 
proposed.  These include: entanglement of 
formation~\cite{Bennett_DiVincenzo_Smolin_Wootters,Wootters98}, 
entanglement of 
distillation~\cite{Bennett_Brassard_Popescu_Schumacher_Smolin_Wootters96}, 
relative entropy of entanglement~\cite{Vedral_Plenio_Jacobs_Knight},  
negativity~\cite{Zyczkowski_Horodecki_Sanpera_Lewenstein98,Vidal_Werner02},
and so on. 
It is worth remarking that even for the smallest Hilbert space capable of 
exhibiting entanglement, i.e.,~the two-qubit system (for which Wootters 
has determined the entanglement of formation~\cite{Wootters98}), there 
are aspects of entanglement that remain to be explored. 

Among the family of mixed quantum mechanical states, special status 
should be accorded to those that, for a given value of the 
entropy~\cite{FOOT:entropy}, have the largest possible degree of 
entanglement~\cite{Munro_James_White_Kwiat01}. The reason for this is that such states can be 
regarded 
as mixed-state generalizations of the Bell states, the latter being known to 
be the maximally entangled 2-qubit pure states.
The notion of maximally entangled mixed states was introduced by 
Ishizaka and Hiroshima~\cite{Ishizaka_Hiroshima00} in a closely 
related setting, i.e., that of 2-qubit mixed states whose entanglement is 
maximized at fixed eigenvalues of the density matrix (rather than at 
fixed entropy of the density matrix).  Evidently, the entanglement of 
the maximally entangled mixed states of Ishizaka and Hiroshima 
cannot be increased by any {\it global} unitary transformation.  For these 
states, it was shown by Verstraete et al.~\cite{Verstraete_Audenaert_deMoor01}
that the maximality property continues to hold if any of the following 
three measures of entanglement---entanglement of formation, negativity, 
and relative entropy of entanglement---is replaced by one of the other two. 

 The question of the ordering of entanglement measures was raised by 
Eisert and Plenio~\cite{Eisert_Plenio99} and investigated numerically 
by them and analytically by Verstraete et al.~\cite{Verstraete_Audenaert_Dehaene_deMoor01}. It was proved by Virmani and Plenio~\cite{Virmani_Plenio00} that all good 
asymptotic entanglement measures are either identical or fail to 
uniformly give consistent orderings of density matrices. This implies that
the resulting maximally entangled mixed states (MEMS) may depend on the
measures one uses to quantify entanglement. Moreover, in finding the form
of MEMS, one needs to quantify the {\em mixedness\/} of a state, and there
can also be ordering problems for mixedness. This implies that the MEMS
may depend on the measures of mixedness as well. 

This Paper is organized as follows. We begin, in Secs.~\ref{SEC:MEMS} 
and \ref{SEC:MeasOfMix}, by reviewing several measures of entanglement 
and mixedness.  In the main part of the Paper, Sec.~\ref{SEC:Frontier}, 
we consider various entanglement-versus-mixedness planes, 
in which entanglement and mixedness are quantified in several ways. 
Our primary objective, then, is to determine the 
{\it frontiers\/}, i.e., the boundaries of the regions occupied by 
physically allowed states in these planes, and to identify the 
structure of 
these maximally entangled mixed states. 
In Sec.~\ref{SEC:Conclude} we make some concluding remarks.
\section{Entanglement Criteria and their measures}
\label{SEC:MEMS}
It is well known that there are a large number of entanglement
measures $E$. For a state described by the density matrix $\rho$
a good entanglement measure must satisfy, at least, the following conditions~\cite{VedralPlenio98,Horodecki300}: \\ 
C1.~(a)~$E(\rho)\ge 0$; 
(b)~$E(\rho)=0$ if $\rho$ is not entangled~\cite{REF:SepEnt}; and 
(c)~$E(\mbox{Bell states})=1$. \\
C2.~For any state $\rho$ and any local unitary transformation, 
i.e., a  unitary transformation of the form $U_A \otimes U_B$, 
the entanglement remains unchanged. \\
C3.~Local operations, classical communication and post\-selection cannot increase the expectation value of the entanglement. \\
C4.~Entanglement is convex under discarding information: 
$\sum_i p_i\,E(\rho_i)\ge E(\sum_i p_i\,\rho_i)$. \\

The entanglement quantities chosen by us satisfy the properties C1--C4. 
Here, we do not impose the condition that any good entanglement measure should
reduce to the {\em entropy of entanglement\/} (to be defined in the following)
for pure states.
\subsection{Entanglement of formation and entanglement cost}
\label{sec:Ef}
The first measure we shall consider is the entanglement of
formation $E_{\rm F}$~\cite{Bennett_DiVincenzo_Smolin_Wootters}; it quantifies the amount of entanglement necessary to
create the entangled state. It is defined by
\begin{eqnarray}
E_{\rm F}(\rho)\equiv\min_{\{p_i,\psi_i\}}\sum_i p_i \, E(|\psi_i\rangle\langle\psi_i|),
\end{eqnarray}
where the minimization is taken over those probabilities $\{p_i\}$ and pure
states $\{\psi_i\}$ that, taken together,
reproduce the density matrix  $\rho=\sum_i p_i
|\psi_i\rangle\langle\psi_i|$. Furthermore,
the quantity $E(|\psi_i\rangle\langle\psi_i|)$ (usually called the {\em entropy
of entanglement})
measures the entanglement of the pure state
$|\psi_i\rangle$ and is defined to be
the von Neumann entropy of the reduced density matrix
$\rho_i^{(A)}\equiv\mbox{Tr}_B |\psi_i\rangle\langle\psi_i|$, i.e.,
\begin{eqnarray}
\label{eqn:EntVon}
E(|\psi_i\rangle\langle\psi_i|)=-\mbox{Tr}\rho_i^{(A)}\,\log_2\rho_i^{(A)}.
\end{eqnarray}

For two-qubit systems, $E_{\rm F}$ can be expressed explicitly as~\cite{Wootters98}
\begin{subequations}
\begin{eqnarray}
E_{\rm F}(\rho)& = & h\left( \frac{1}{2}[1 + \sqrt{1-C(\rho)^2}] \right),
\label{eqn:Wootters} \\
h(x) & \equiv & -x \log_2 x - (1-x) \log_2(1-x),
\label{eqn:entropy}
\end{eqnarray}
where $C(\rho)$, the {\em concurrence} of the state $\rho$, is defined as
\begin{eqnarray}
        C(\rho) \equiv \max \{0,\sqrt{\lambda_1}-\sqrt{\lambda_2}-
 \sqrt{\lambda_3}-\sqrt{\lambda_4}\},
\end{eqnarray}%%
\end{subequations}
in which 
$\lambda_1,\ldots,\lambda_4$ are the eigenvalues of the matrix
$\rho (\sigma_y\otimes\sigma_y)\rho^{*}(\sigma_y\otimes\sigma_y)$
in nondecreasing order and $\sigma_y$ is a Pauli spin matrix.
$E_{\rm F}(\rho)$, $C(\rho)$, and the {\it tangle\/}
$\tau(\rho)\equiv C(\rho)^2$ are equivalent measures of entanglement,
inasmuch as they are monotonic functions of one another.

A measure associated with the entanglement of formation is the entanglement 
cost $E_{\rm C}$~\cite{Bennett_DiVincenzo_Smolin_Wootters} which is defined 
via 
\begin{eqnarray}
E_{\rm C}(\rho)\equiv \lim_{n\rightarrow\infty}\frac{E_{\rm F}(\rho^{\otimes n})}{n}.
\end{eqnarray}
This is the asymptotic value of the average entanglement of formation. $E_{\rm C}$ is, in general, difficult to calculate.

\subsection{Entanglement of distillation 
and relative entropy of entanglement}

Related to the entanglement of formation is the entanglement of
distillation $E_{\rm D}$ \cite{Bennett_Brassard_Popescu_Schumacher_Smolin_Wootters96}, 
which characterizes the amount of entanglement
of a state $\rho$ as the fraction of Bell states that
can be distilled using the optimal purification procedure:
$E_{\rm D}(\rho)\equiv\lim_{n\rightarrow\infty} m/n$, 
where $n$ is the number of copies of $\rho$ used and $m$ is the
maximal number of Bell states that can be distilled from them.
The difference $E_{\rm F}-E_{\rm D}$ can be regarded as {\em undistillable
entanglement\/}.
$E_{\rm D}$ is a difficult quantity to calculate, but
the relative entropy of entanglement 
$E_{\rm R}$~\cite{Vedral_Plenio_Jacobs_Knight},
which we shall define shortly, provides an upper bound on $E_{\rm D}$ and is more readily calculable than it. 
For this reason, it is the second measure that we
consider in this Paper. It is defined variationally via 
\begin{equation}
E_{\rm R}(\rho)\equiv\min_{\sigma\in D}{\rm
Tr}\left(\rho\log\rho-\rho\log\sigma\right),
\end{equation}
where $D$ represents the (convex) set of all separable 
density operators $\sigma$. In certain ways, the relative 
entropy of entanglement can be viewed as a {\it distance\/} ${\cal D}(\rho||\sigma^\ast)$ 
from the entangled state $\rho$ to the closest separable 
state $\sigma^\ast$.
We remark that for pure states of two-qubit systems the relative 
entropy has the same value as the entanglement of formation.

\subsection{Negativity}
The third measure that we shall consider is  the {\it negativity\/}.
The concept of the negativity of a state is closely related to the
well-known Peres-Horodecki condition for the separability of a state
\cite{Peres96_Horodecki396}. If a state is separable (i.e., not 
entangled) then the partial transpose of its density matrix is 
again a valid state, i.e., it is positive semi-definite.  
It turns out that the partial transpose of a non-separable
state has one or more negative eigenvalues. The negativity of a
state \cite{Zyczkowski_Horodecki_Sanpera_Lewenstein98} indicates the 
extent to which a state violates the positive partial transpose separability criterion. We will adopt the definition of
negativity as twice the absolute value of the sum of the negative
eigenvalues:
\begin{equation}
N(\rho) = 2\max(0,-\lambda_{\rm neg}),
\end{equation}
where $\lambda_{\rm neg}$ is the sum of the negative eigenvalues of 
$\rho^{{\rm T}_B}$.  In $C^2\otimes C^2$ (i.e., two-qbit) systems  
it can be shown that the partial transpose of the density matrix can 
have at most one negative eigenvalue (see App.~\ref{app:OneNeg}). 
It was proved by Vidal and Werner~\cite{Vidal_Werner02} that 
negativity is an entanglement {\em monotone}, i.e., it satisfies
criteria C1--C4 and, hence, is a good entanglement measure. 
We remark that for two-qubit pure states the negativity gives the 
same value as the concurrence does.

\subsection{Bures metric}
The Bures metric of entanglement is defined as
\begin{eqnarray}
E_{\rm B}(\rho)\equiv 
\min_{\sigma\in D}\big(2-2\sqrt{F(\rho,\sigma)}\big),
\end{eqnarray}% 
where $F(\rho,\sigma)\!\equiv\! 
\big({\rm Tr}\sqrt{\sqrt{\sigma}\rho\sqrt{\sigma}}\big)^2$ 
is the {\it fidelity\/}.  In the same way that relative entropy can, 
this entanglement measure can be viewed as the distance from the 
closest separable state to the entangled state considered, where the 
distance is now defined by 
${\cal D}(\rho||\sigma)\equiv
\big(2-2\sqrt{F(\rho,\sigma)}\big)$~\cite{VedralPlenio98}. 
We remark that for two-qubit pure states the Bures metric reduces to 
the tangle defined in Sec.~\ref{sec:Ef}.

\subsection{Lewenstein-Sanpera entanglement}
It was shown by Lewenstein and Sanpera~\cite{Lewenstein_Sanpera98} that
any density matrix $\rho$ has a decomposition into two parts:
\begin{eqnarray}
\label{eqn:LSdec}
\rho=\lambda \rho_{\rm s} + (1-\lambda) \rho_{\rm e},
\end{eqnarray}
where $\rho_{\rm s}$ is separable, $\rho_{\rm e}$ is entangled, and the 
weight $\lambda$ is maximal, in which case the decomposition is unique. 
They refer to $\rho_{\rm s}$ as the {\it best separable approximation\/} 
(BSA) to $\rho$. It should be pointed out that, in general, it is 
nontrivial to establish the decomposition, even in the simplest relevant setting of $C^2\otimes C^2$ systems. Evidently, $\lambda$ and $\rho_{\rm e}$ 
contain information about the entanglement of $\rho$. Karnas and Lewenstein~\cite{Karnas_Lewenstein01} later showed that the quantity $E_{\rm LS}\equiv(1-\lambda)$, which we will call {\em the LS entanglement}, satifies the above criteria, and hence is a good entanglement measure. 
For $C^2\otimes C^2$ systems it turns out that $\rho_e$ is a pure state, i.e., $|\psi_{\rm 
e}\rangle\langle\psi_{\rm e}|$, and in this case it suggests that the quantity $\tilde{E}_{\rm LS}$, defined via
\begin{eqnarray}
\tilde{E}_{\rm LS}\equiv
(1-\lambda)E(|\psi_{\rm e}\rangle\langle|\psi_{\rm e}|),
\label{eqn:LS}
\end{eqnarray}
may also be a good entanglement measure.  
We remark that the for two-qubit case the entanglement measure 
($1-\lambda$) is known to be equal to the Schmidt measure introduced in 
Ref.~\cite{Eisert_Briegel01}.

Even though the LS decomposition is not, in general, straightforward to 
find for the states in Eq.~(\ref{eqn:Ansatz}), the LS decomposition reads
\begin{equation}
\label{eqn:LSDecomp}
\begin{pmatrix}
              x\!+\!\frac{\mygamma}{2}\!&\! 0 \!&\! 0 \!&\! \frac{\mygamma}{2} \cr
              0\!  &\! a \!&\! 0 \!&\! 0 \cr
              0\!  & \!0 \!&\! b\! &\! 0 \cr
              \frac{\mygamma}{2} \!\!&\! 0 \!&\! 0 \!&\!
	       y\!+\!\frac{\mygamma}{2} \cr
            \end{pmatrix}\!\!= 
	      \!\!
    \begin{pmatrix}
              x\!+\!\sqrt{ab} \!&\! 0 \!&\! 0 \!&\! \sqrt{ab} \cr
              0  \!&\! a \!&\! 0 \!&\! 0 \!\cr
              0  \!&\! 0 \!&\! b \!&\! 0 \!\cr
              \sqrt{ab} \!&\! 0 \!&\! 0 \!&\! y\!+\!\sqrt{ab} \cr
            \end{pmatrix}\!+\!(1\!-\!\lambda)\rho_e,
\end{equation}
where 
$\rho_{\rm e}=|\phi^+\rangle\langle\phi^+|$ 
(with 
$|\phi^\pm\rangle\equiv(|00\rangle\!\pm\!|11\rangle)/\sqrt{2}$), $(1-\lambda)=\mygamma-2\sqrt{ab}$, 
and we consider only $\mygamma-2\sqrt{ab}>0$ 
(as for 
$\mygamma-2\sqrt{ab}\le 0$ 
the whole density matrix is separable).

If we compute the concurrence of the state in Eq.~(\ref{eqn:Ansatz}), 
we find $C=\max\{0,\mygamma-2\sqrt{ab}\}$.  It is interesting to note 
that the LS entanglement gives, for the Ansatz states in 
Eq.~(\ref{eqn:Ansatz}), the same value as the concurrence does;
but for an arbitrary state this is not, in general, true~\cite{WellensKus01}. 

\subsection{Ordering difficulties with entanglement measures}
\begin{figure} 
\vspace{0.5cm}
\centerline{
\rotatebox{0}{
	\epsfxsize=7cm
	\epsfbox{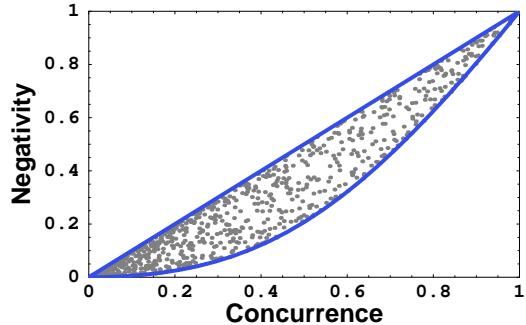}
}
}
%\centerline{$S$}
\vspace{-0.2cm}
\caption{Comparison of negativity and concurrence. Dots represent randomly generated states. (The form of the boundary is discussed in the text.) It is
apparent that these two measures of entanglement can give
different orderings for pairs of states.}
\label{Fig:CvsN}
\end{figure}

We now pause to touch on certain difficulties posed by the task of 
ordering physical states using entanglement. As first discussed and 
explored numerically by Eisert and Plenio~\cite{Eisert_Plenio99}, 
and subsequently investigated analytically by Verstraete et al.~\cite{Verstraete_Audenaert_Dehaene_deMoor01}, 
different entanglement measures can give different orderings for 
pairs of mixed states.  This can be seen, e.g., from the plot of 
concurrence versus negativity, Fig.~\ref{Fig:CvsN}.  The upper 
boundary is readily seen to be $N\le C$ whereas the lower boundary 
can be derived, giving 
$N\ge \sqrt{2(C\!-\!\frac{1}{2})^2\!+\!\frac{1}{2}}+(C\!-\!1)$; see 
Ref.~\cite{Verstraete_Audenaert_Dehaene_deMoor01}.
Hence, when we wish to explain maximally entangled mixed states  
we need to be very explicit about the measure of entanglement (and also 
mixedness; see the following section). Different measures are likely to 
lead to different classes of MEMS states. 

We end this section by mentioning the three entanglement measures that 
we shall use to compute the entanglement-mixedness frontiers: 
entanglement of formation, 
negativity, and 
relative entropy of entanglement.  
The first two of these are straightforward to compute, at least in 
two-qubit settings.  For the third, certain results are available~\cite{VedralPlenio98,Verstraete_Audenaert_deMoor01} that 
ease the computation of MEMS.  We have also reviewed four additional 
measures (entanglement cost, entanglement of distillation, LS 
entanglement, and the Bures metric). Of these, however, the first 
two are rather difficult to compute, let alone maximize; the third 
is also difficult to compute, at least in practice. As for the 
fourth, calculating the entanglement involves finding the closest 
separable states (as is required for the case of relative entropy).   

\section{Measures of mixedness}
\label{SEC:MeasOfMix}
In the entanglement-measure literature, 
two measures of mixedness have basically 
%** (basically?)
been used: 
$\big(1-{\rm Tr}[\rho^2]\big)$ 
and the von Neumann entropy. 
Whereas the latter has an natural significance stemming from 
its connections with statistical physics and information theory, 
the former is substantially easier to calculate.  Of course, for 
density matrices that are almost completely mixed, the two 
measures show the same trend. 
   
\subsection{von Neumann entropy}
The von Neumann entropy, the standard measure of randomness of a 
statistical ensemble described by a density matrix, is defined by
\begin{eqnarray}
S_{\rm V}(\rho)\equiv
-\mathrm{Tr}(\rho\log\rho)
= -\sum_i \lambda_i \,\log \lambda_i\,,
\label{eqn:vonNeumann}
\end{eqnarray}
where $\lambda_i$ are the eigenvalues of the density matrix 
$\rho$ and the $\log$ is taken to base ${\cal N}$, the 
dimension of the Hilbert space in question.
It is straightforward to show that the extremal values of 
$S_{\rm V}$ are zero (for pure states) and unity (for 
completely mixed states). To compute the von Neumann entropy 
it is necessary to have the full knowledge of the eigenvalue 
spectrum.

As we shall mention in the following subsection, there is a linear entropy threshold above which all states are separable. Qualitatively identical behavior is encountered for the von Neumann entropy. In particular, as we shall see in Sec.~\ref{sec:EfSv}, for two-qubit systems all states
are separable for $S_{\rm V}\ge -(1/2)\log_4(1/12)\approx 0.896$.

\subsection{Purity and linear entropy}
The second measure that we shall consider is called the 
{\it linear entropy\/} and is based on the purity of a state, 
${\cal P}\equiv{\rm Tr}\left[\rho^2\right]$, 
which ranges from 1 (for a pure state) to $1/{\cal N}$ for a 
completely mixed state with dimension ${\cal N}$. 
The linear entropy $S_{\rm L}$ is defined via 
\begin{eqnarray}
S_{\rm L}(\rho)\equiv 
\frac{{\cal N}}{{\cal N}-1}(1-\mbox{Tr}[\rho^2]), 
\label{eqn:LinearEntropy}
\end{eqnarray} 
which ranges from 0 (for a pure state) to 1 (for a maximally mixed
state). The linear entropy is generally a simpler quantity to
calculate than the von Neumann entropy as there is no need for 
diagonalization.  For $C^2\otimes C^2$ systems the linear entropy 
can be written explicity as 
\begin{eqnarray}
S_{\rm L}(\rho)\equiv 
\frac{4}{3}(1-\mbox{Tr}\left[\rho^2\right]).
\end{eqnarray} 

A related measure, which we shall not use in this Paper (but mention 
for the sake of completeness), is {\it the inverse participation ratio\/}.   
Defined via  $R\equiv 1/ \mbox{Tr}\left[\rho^2\right]$, 
it ranges from $1$ (for a pure state) to ${\cal N}$ 
(for the maximally mixed state).  
An attractive property of the inverse participation ratio is that 
% ** it is known that 
all states with $R\geq {\cal N}-1$ are separable~\cite{Zyczkowski_Horodecki_Sanpera_Lewenstein98}, 
which implies all states with a linear entropy
$S_{\rm L}(\rho)\geq {\cal N}{({\cal N} -2)}/\left({\cal N}-1\right)^2$ 
(which is $8/9$ when ${\cal N}=4$) are separable.

\subsection{Comparing linear and von Neumann entropies}
\label{sec:SlSv}
The aim of this subsection is to illustrate the difference between the 
linear and von Neumann entropies. We shall do this by considering the 
${\cal N}=4$ Hilbert space, and seeking the highest and lowest von Neumann 
entropies consistent with a given value of linear entropy.  Before 
restricting ${\cal N}$ to 4, the corresponding stationarity problem reads: 
\begin{eqnarray}
\delta \left( S_{\rm V}(\rho) 
+\beta\frac{{\cal N}-1}{2{\cal N}} 
S_{\rm L}(\rho)-(\nu-1)\mbox{Tr}\rho \right) =0,
\end{eqnarray}
where $\beta$ and $\nu$ are, respectively, Langrange multipliers that 
enforce the constraints that linear entropy be fixed and that $\rho$ 
be normalized. Thus, we arrive at the engaging self-consistency condition
\begin{eqnarray}
\label{eqn:Stationarity}
\rho=\exp (-\nu-\beta \rho),
\end{eqnarray}
in which $\nu$ and $\beta$ can be fixed upon implementing the constraints.

\begin{figure} 
\vspace{0.5cm}
\centerline{
\rotatebox{0}{
	\epsfxsize=7cm
	\epsfbox{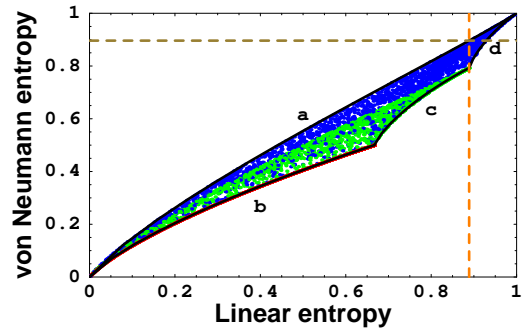}
}}
%\centerline{$S$}
\vspace{-0.2cm}
\caption{Comparison of linear entropy and von Neumann entropy. Dots represent randomly generated states; pure (Rank-1) states lie at the origin; Rank-2 states lie on segment $b$; the lighter dots in the interior are Rank-3 states; the darker ones are Rank-4 states. The lower boundary comprises three segments meeting at cusps, whereas the upper boundary is a smooth curve. The two dashed lines represent thresholds of entropies beyond which no states contain entanglement. 
}
\label{Fig:SvSl}
\end{figure}
By hypothesizing certain relationships among the eigenvalues of the density matrix, we have 
been able, for the case of ${\cal N}=4$, to find certain solutions of the self-consistency 
condition and, hence, a candidate for the boundary of the region of $S_{\rm L}$ vs.~$S_{\rm V}$ 
plane (see Fig.~\ref{Fig:SvSl}) corresponding to legitimate density matrices. These solutions, 
when given in terms of eigenvalues, read
\begin{subequations}
\begin{eqnarray}
&&\Big\{\frac{1\!-\!\mygamma}{4}\!,\frac{1\!-\!\mygamma}{4}\!,\frac{1\!-\!\mygamma}{4}\!,
\frac{1\!+\!3\mygamma}{4}\!\Big\},
\ \mbox{for} \ 0\le\mygamma\le 1, \\
&&\{\mygamma,1-\mygamma,0,0\}, \qquad \qquad\quad \ \mbox{for} \  \frac{1}{2}\le\mygamma\le 1, \\
&&\Big\{\mygamma,\frac{1\!-\!\mygamma}{2},\frac{1\!-\!\mygamma}{2},0\Big\}, \  \qquad\quad \mbox{for}  \ 
0\le\mygamma\le\frac{1}{3}, \\  
&&\Big\{\frac{4\!-\!\mygamma}{12},\frac{4\!-\!\mygamma}{12},\frac{4\!-\!\mygamma}{12},
\frac{3\mygamma}{12}\Big\}, \ \ \  \mbox{for} \ 0\le\mygamma\le 1,
\end{eqnarray}
\label{eqn:SvSlSolutions}%%
\end{subequations}
and they correspond to the upper boundary, and the lowest, middle, and highest pieces of the 
lower boundary, respectively. Note that the lower boundary comprises three (in general, ${\cal N}-1$) segments that meet at cusps. We remark, parenthetically, that the solutions with zero 
eigenvalues correspond to extrema within some subspace spanned by those eigenvectors with nonzero eigenvalues, and therefore only obey the stationarity 
condition~(\ref{eqn:Stationarity}) within the subspace. 

Is there any significance to the boundary states? 
Boundary segment~(a) includes the Werner states defined in Eq.~(\ref{eqn:Werner}). 
Boundary segment~(b) includes the first branch of the MEMS for $E_{\rm F}$ and 
$S_{\rm L}$ specified below in Eq.~(\ref{eqn:MEMS}).
% ** (which we denote symbolically by 
% 	$\{{\rm MEMS}:E_{\rm F},S_{\rm L}\}$)
The segment~(c) includes the states 
\begin{eqnarray}
\label{eqn:Gisin}
\rho_c= \mygamma
|\phi^+\rangle
\langle\phi^+|
+\frac{1\!-\!\mygamma}{2}\big(|01\rangle\langle01|\!
+\!|10\rangle\langle10|\big).
\end{eqnarray}%% 
States on segment~(d) are all unentangled.  Of course, the boundary 
segments include not only the specified states but also all states 
derivable from them by global unitary transformation.

As for the interior, we have obtained this numerically by constructing 
a large number of random sets of eigenvalues of legitimate density 
matrices, and computing for each the two entropies.  As Fig.~\ref{Fig:SvSl} 
shows, no points lie outside the boundary curve, providing confirmatory 
evidence for the forms given in Eq.~(\ref{eqn:SvSlSolutions}).
 
The fact that the bounded region is two-dimensional indicates the lack 
of precision with which the linear entropy characterizes the von Neumann 
entropy (and vice versa, if one wishes). In particular, the figure reveals 
an ordering difficulty: pairs of states, $A$ and $B$, exist for which 
$S_{\rm L}^A-S_{\rm L}^B$ and $S_{\rm V}^A-S_{\rm V}^B$ differ in sign.  
Worse still, states having a common value of $S_{\rm V}$ have a continuum 
of values of $S_{\rm L}$, and vice versa. 
% ** (Do we really mean worse still?)

\section{Entanglement-versus-mixedness frontiers}
\label{SEC:Frontier}
We now attempt to identify regions in the plane spanned by entanglement 
and mixedness that are inhabited by physical states (i.e., characterized 
by legitimate density matrices).  We shall consider the various measures 
of entanglement and mixedness discussed in the previous section.  Of 
particular interest will be the structure of the states that inhabit the 
{\em frontier\/}, i.e., the boundary delimiting the region of physical 
states. Frontier states are maximal in the following sense: for a given 
value of mixedness they are maximally entangled; for a given value of 
entanglement they are maximally mixed.  

\subsection{Parametrization of maximal states}
\label{sec:Rhomems}
The aim of this subsection is to derive the general form of the maximal 
states given in Eq.~(\ref{eqn:Rhomems}), which is what we will use to 
parametrize maximal states.  In Ref.~\cite{Verstraete_Audenaert_deMoor01}, 
it is shown that, given a fixed set of eigenvalues, all states that 
maximize one of the three entanglement measures 
(entanglement of formation, negativity or relative entropy) 
automatically maximize the other two.  It was further shown that the global 
unitary transformation that takes arbitrary states into maximal ones has 
the form
\begin{eqnarray}
U=(U_1\otimes U_2)T D_\phi \Phi^\dag,
\end{eqnarray}
where $U_1$ and $U_2$ are arbitary local unitary transformations, 
\begin{eqnarray}
T\equiv \begin{pmatrix}
                  0 & 0 & 0 & 1 \cr
		  \frac{1}{\sqrt{2}} & 0 & \frac{1}{\sqrt{2}} & 0  \cr
		  \frac{1}{\sqrt{2}} & 0 & \frac{-1}{\sqrt{2}} & 0 \cr
		  0 & 1 & 0 & 0 \cr
		  \end{pmatrix},
\end{eqnarray}
$D_\phi$ is a
unitary diagonal matrix, and $\Phi$ is the unitary matrix that diagonalizes
the density matrix $\rho$, i.e., $\rho=\Phi \Lambda \Phi^\dag$,
where $\Lambda$ is a diagonal matrix, the diagonal elements of which are the four
eigenvalues of $\rho$ listed in nonincreasing order ($\lambda_1
\ge\lambda_2\ge\lambda_3\ge\lambda_4$). Hence, the general form
of a density matrix that is maximal, given a set of eigenvalues, is (up to local unitary transformations)
\begin{equation}
T \begin{pmatrix}
		  \lambda_1 & 0 & 0 & 0 \cr
		  0 & \lambda_2 & 0 & 0 \cr
		  0 & 0 &\lambda_3& 0 \cr
		  0 & 0 & 0 & \lambda_4 \cr
		  \end{pmatrix} T^\dagger
     \!=\!\begin{pmatrix}
                \lambda_4 \!&\! 0 \!&\! 0 \!&\! 0 \cr
		0 \!&\! \frac{\lambda_1\!+\!\lambda_3}{2} \!&\!
		\frac{\lambda_1\!-\!\lambda_3}{2} \!&\! 0 \cr
                0 \!&\! \frac{\lambda_1\!-\!\lambda_3}{2} \!&\!
		\frac{\lambda_1\!+\!\lambda_3}{2} \!&\! 0 \cr
		0 \!&\! 0 \!&\! 0 \!&\! \lambda_2 \cr
            	\end{pmatrix}\!.		  
\end{equation}
This matrix is locally equivalent to the form 
\begin{equation}
\label{eqn:Rhomems}
%\rho_{{\rm mems}}=
\begin{pmatrix}
  x+\frac{\mygamma}{2} & 0 & 0 & \frac{\mygamma}{2} \cr
              0  & a & 0 & 0 \cr
              0  & 0 & b & 0 \cr
              \frac{\mygamma}{2} & 0 & 0 & x+\frac{\mygamma}{2}\cr
	      \end{pmatrix},
\end{equation}%%
with $x\!+\!\frac{\mygamma}{2}\!=\!(\lambda_1\!+\!\lambda_3)/2$, $\mygamma\!=\!\lambda_1\!-\!\lambda_3$, $a\!=\!\lambda_2$, and $b\!=\!\lambda_4$.
The above derivation justifies the Ansatz form~(\ref{eqn:Ansatz}) used in Ref.~\cite{Munro_James_White_Kwiat01} to derive the entanglement of formation vs.~linear entropy MEMS. We remark that one may as well use the four eigenvalues
$\lambda_i$'s as the parametrization. Nevertheless, the form~(\ref{eqn:Rhomems}), as well as (\ref{eqn:Ansatz}),
can be nicely viewed as a mixture of a Bell state $|\phi^+\rangle$ with some diagonal separable mixed state. 

\subsection{Entanglement-versus-linear-entropy frontiers}
We begin by measuring mixedness in terms of the linear entropy, 
and comparing the frontier states for various measures of entanglement.

\subsubsection{Entanglement of formation}
The characterization of physical states in terms of their entanglement 
of formation and linear entropy was introduced by Munro et al.~in 
Ref.~\cite{Munro_James_White_Kwiat01}. (Strictly speaking, they 
considered the tangle rather than the equivalent entanglement of 
formation.)\thinspace\ 
Here, we shall consider yet another equivalent quantity: concurrence 
(see Sec.~\ref{sec:Ef}).  
In order to find the frontier, Munro et al.~proposed Ansatz states of the form
\begin{eqnarray}
\label{eqn:Ansatz}
\rho_{\mbox{\tiny Ansatz}}=\begin{pmatrix}
	     x+\frac{\mygamma}{2} & 0 & 0 & \frac{\mygamma}{2} \cr
              0  & a & 0 & 0 \cr
              0  & 0 & b & 0 \cr
              \frac{\mygamma}{2} & 0 & 0 & y+\frac{\mygamma}{2} \cr 
	      \end{pmatrix}, 
\end{eqnarray}
where $x,y,a,b,\mygamma\ge 0$ and $x+y+a+b+\mygamma=1$.
They found that, of these, the subset 
\begin{subequations}
\begin{eqnarray}%\label{eqn:MEMSSS}
\rho_{ \mbox{\tiny MEMS:$E_{\rm F},S_{\rm L}$}}=
\begin{cases} \ \rho_{\rm I}(\mygamma), &{\rm for \ }\frac{2}{3}\le\mygamma\le 1;\cr
              \ \rho_{\rm I\!I}(\mygamma), &{\rm for\ } 0\le\mygamma\le\frac{2}{3};\cr
	    \end{cases}\quad\quad&& \\
\rho_{\rm I}(\mygamma)\!=\!\begin{pmatrix}\frac{\mygamma}{2} \!&\! 0 \!&\! 0 & \frac{\mygamma}{2} \cr
             0  \!&\! 1\!\!-\!\!\mygamma \!&\! 0 & 0 \cr
             0  \!&\! 0 \!&\! 0 & 0 \cr
             \frac{\mygamma}{2} \!&\! 0 \!&\! 0 & \frac{\mygamma}{2}\cr
	     \end{pmatrix}\!, \ \rho_{\rm I\!I}(\mygamma)\!=\!\begin{pmatrix}
 \frac{1}{3} & 0 & 0 & \frac{\mygamma}{2} \cr
             0  & \frac{1}{3}& 0 & 0 \cr
             0  & 0 & 0 & 0 \cr
	     \frac{\mygamma}{2} & 0 & 0 & \frac{1}{3}\cr
	     \end{pmatrix}\!,&&
\end{eqnarray}%%
\label{eqn:MEMS}%%
\end{subequations}%%
lies on the boundary in the tangle vs.~linear-entropy plane and, 
accordingly,  named these MEMS, in the sense that these states have 
maximal tangle for a given linear entropy.  We remark that at the 
crossing point of the two branches, $\mygamma=2/3$, the density 
matrices on either side coincide.  
\begin{figure}
\vspace{0.5cm}
\centerline{
\rotatebox{0}{
	\epsfxsize=6cm
	\epsfbox{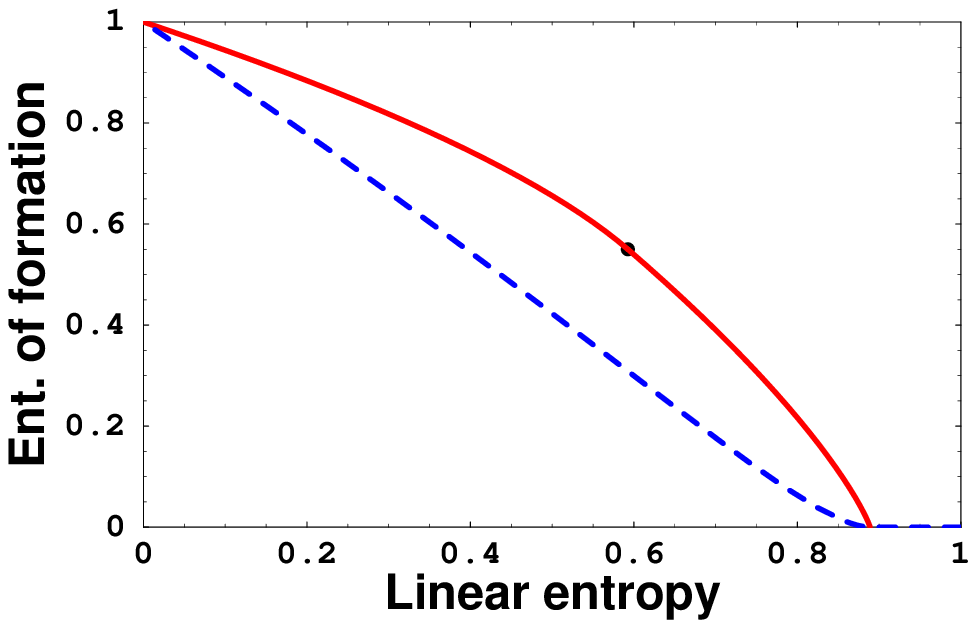}
}}
\vspace{-0.1cm}
\centerline{
\rotatebox{0}{
	\epsfxsize=6cm
	\epsfbox{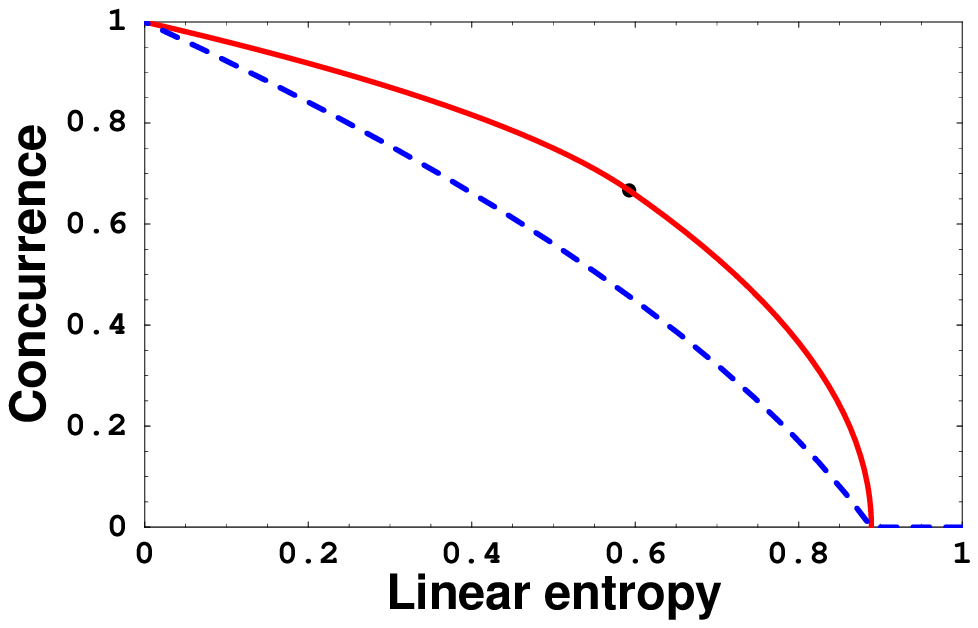}
}}
%\centerline{$E_F$}
\vspace{-0.2cm}
\caption{Entanglement frontier. 
Upper panel: entanglement of formation vs.~linear entropy. 
Lower panel: concurrence vs.~linear entropy.  
The states on the boundary (solid curve) are 
$\rho_{\mathrm{MEMS};E_{\rm F},S_{\rm L}}$.  
A dot indicates a transition from one branch of MEMS to another.  
The dashed curve below the boundary contains Werner states.}
\label{Fig:CSl}
\end{figure}

In Fig.~\ref{Fig:CSl} we plot the entanglement of 
formation/concurrence vs.~linear entropy for the 
family of MEMS~(\ref{eqn:MEMS}); this gives the frontier curve.  For the 
sake of comparison, we also give the curve associated with the family of 
Werner states of the form
\begin{equation}
\rho_{\rm W}  \equiv\mygamma|\phi^+\rangle\langle\phi^+|\!+\!\frac{1-\mygamma}{4}\openone=
       \begin{pmatrix}
       \frac{1+\mygamma}{4} & 0 & 0 & \frac{\mygamma}{2} \cr
               0  & \frac{1-\mygamma}{4} & 0 & 0 \cr
               0  & 0 &  \frac{1-\mygamma}{4} & 0 \cr
               \frac{\mygamma}{2} & 0 & 0 & \frac{1+\mygamma}{4}
	       \end{pmatrix}. 
\label{eqn:Werner}
\end{equation}
Evidently, for a given value of linear entropy these MEMS 
(which we shall denote by 
$\{{\rm MEMS}\!:\!E_{\rm F},S_{\rm L}\}$) 
achieve the highest concurrence.  As the tangle $\tau$ and entanglement 
of formation $E_{\rm F}$ are monotonic functions of the concurrence, Eq.~(\ref{eqn:MEMS}) also gives the boundary curve for these measures.  
This raises an interesting question: Is (\ref{eqn:MEMS}) optimal for 
other measures of entanglement?

\subsubsection{Relative entropy as the entanglement measure}
To find the frontier states for the relative entropy of entanglement we
again turn our attention to the maximal density matrix (\ref{eqn:Rhomems}). 
For this form of density matrix the linear entropy is given 
(with $x$ expressed in terms of $a,b,\mygamma$) by
\begin{eqnarray}
S_{\rm L}=\frac{2}{3}\!\left( -3 a^2 + 2 a \left(1 \!-\! b \right)  +
      \left(1 \!-\! b \right) \left( 1 \!+\! 3 b \right)
      \!-\!\mygamma^2 \right)\!.
\end{eqnarray}
To calculate the relative entropy of entanglement, we need to determine the 
closest separable state to (\ref{eqn:Rhomems}).  It is simpler to do this 
analysis via several cases. We begin by considering the Rank-2 and Rank-3 
cases of~(\ref{eqn:Rhomems}). We set $b=0$ ($\lambda_4=0$) and express $x$ 
in terms of $a$ and $\mygamma$
in the density matrix, obtaining
\begin{eqnarray}
\label{eqn:rank2n3}
\rho=\begin{pmatrix}
              \frac{1-a}{2} & 0 & 0 & \frac{\mygamma}{2} \cr
              0  & a & 0 & 0 \cr
              0  & 0 & 0 & 0 \cr
              \frac{\mygamma}{2} & 0 & 0 & \frac{1-a}{2} 
            \end{pmatrix},
\end{eqnarray}
and thus find that the closest 
separable density matrix $\sigma^*$ is given by~\cite{VedralPlenio98}
\begin{subequations}
\begin{eqnarray}
\sigma^\ast&=&\begin{pmatrix}
                  C & 0 & 0 & D \cr
		  0 & E  & 0 & 0 \cr
		  0 & 0 & 1\!-\!2 C\!-\!E & 0 \cr
		  D & 0 & 0 & C\cr
		  \end{pmatrix}, \\
 C&\equiv&\frac{\left( 1\! +\! a \right) \left( 1 \!-\! a^2 \!-\! \mygamma^2 \right) }
 {2 \left( 1\! +\! a \!-\! \mygamma \right) \left( 1 \!+\! a \!+\! \mygamma \right) }, \\
 D&\equiv&\frac{a\left( 1 \!+\! a \right)\mygamma}
     {\left( 1\! +\! a \!-\! \mygamma \right)\left( 1 \!+\! a \!+\! \mygamma \right)}, \\  
E&\equiv&\frac{a\left( 1 \!+\! a \right)^2}
  {\left( 1 \!+\! a \!-\! \mygamma \right)\left( 1\! +\! a \!+\! \mygamma \right) }.
\end{eqnarray}  
\end{subequations}
The relative entropy of entanglement is now simply given by 
\begin{eqnarray}
\label{eqn:ER3}
&&E_{\rm R}(\rho)={\rm Tr}\left(\rho\log\rho-\rho\log\sigma^\ast\right) \nonumber\\
&& \quad=\!\frac{1\!+\!a}{2}\log\frac{(1\!+\!a)^2\!-\!\mygamma^2}{(1\!+\!a)^2}+\frac{\mygamma}{2}
\log\frac{1\!+\!a\!+\!\mygamma}{1\!+\!a\!-\!\mygamma},
\end{eqnarray}
with the linear entropy being given by 
\begin{eqnarray}
\label{eqn:SL3}
S_{\rm L}\!=\!\frac{2}{3}\! \left(1\!+\!2a\!-\!3a^2-\!\mygamma^2\right),
\end{eqnarray}
subject to the constraint $(a+\mygamma)\le 1$. 
For the Rank-2 case $a=1-\mygamma$ ($b=x=0$), and the resulting solution is the
Rank-2 matrix $\rho_{\rm I}(\mygamma)$ given in Eq.~(\ref{eqn:MEMS}) with $1/2\le\mygamma\le 1$. We remark that this Rank-2 solution is always a candidate MEMS for the three entanglement measures that we consider in this Paper. In order to determine whether
or in what range the Rank-2 solution achieves the global maximum, we need to compare it with the Rank-3 and Rank-4 solutions.

By maximizing $E_{\rm R}(\rho)$ for a given value of $S_{\rm L}$, we find the
following stationary condition:
\begin{eqnarray}
\label{eqn:stationaryReSL}
\mygamma\log\frac{(1+a)^2-\mygamma^2}{(1+a)^2}=(3a-1)\log\frac{1+a+\mygamma}{1+a-\mygamma}.
\end{eqnarray}
Given a value of $S_{\rm L}$, we can solve Eqs.~(\ref{eqn:SL3}) and (\ref{eqn:stationaryReSL}), at least numerically, to obtain the parameters $a$ and $\mygamma$, and hence, from Eq.~(\ref{eqn:rank2n3}), the Rank-3 MEMS. However,
if the constraint inequality $a+\mygamma\le 1$ turns out to be violated, the solution is invalid. 

We now turn to the Rank-4 case. It is straightforward, if tedious, to show 
that the Werner states, Eq.~(\ref{eqn:Werner}), obey the stationarity 
conditions appropriate for Rank~4.  However, it turns out that this solution 
is not maximal.
\begin{figure}
\vspace{0.5cm}
\centerline{
\rotatebox{0}{
	\epsfxsize=7cm
	\epsfbox{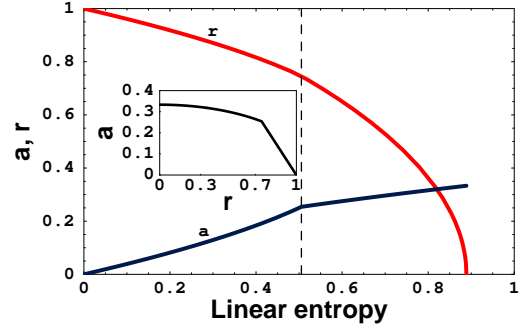}}}
\vspace{-0.2cm}
\caption{Dependence of $a$ and $\mygamma$ of the frontier states on 
linear entropy.  The dotted line indicates the transition between two 
branches of MEMS.}
\label{Fig:ARSl}
\end{figure}

\begin{figure}
\vspace{0.5cm}
\centerline{
\rotatebox{0}{
	\epsfxsize=7cm
	\epsfbox{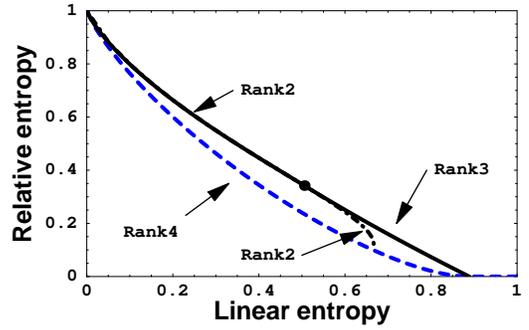}}}
\vspace{-0.2cm}
\caption{Entanglement frontier: 
relative entropy of entanglement vs.~linear entropy.  
The frontier states (solid curve) are 
$\rho_{\mathrm{MEMS};E_{\rm R},S_{\rm L}}$. 
The dot indicates the transition between branches of MEMS.}
\label{Fig:RSl}
\end{figure}

To summarize, the frontier states, which we denote by 
$\{{\rm MEMS}\!:\!E_{\rm R},S_{\rm L}\}$, 
are states of the form~(\ref{eqn:rank2n3});  the dependence of the 
parameters $a$ and $\mygamma$ on $S_{\rm L}$ is shown in 
Fig.~\ref{Fig:ARSl}.  In Fig.~\ref{Fig:RSl}, we show the resulting 
frontier, as well as curves corresponding to non-maximal stationary 
states. The frontier states have the following structure: 
(i)~for $S_{\rm L} \alt 0.5054$ they are the Rank-2 MEMS of 
Eq.~(\ref{eqn:MEMS}) but with $\mygamma$ restricted to the range 
from $1$ (at $S_{\rm L}=0$) to approximately $0.7459$ 
(at $S_{\rm L}\simeq 0.5054$); 
(ii)~for $S_{\rm L} \agt 0.5054$ the MEMS are Rank~3, with 
parameters $a$ and $\mygamma$ satisfying Eqs.~(\ref{eqn:stationaryReSL}) 
and (\ref{eqn:SL3}) at each value of $S_{\rm L}$, and $(a,\mygamma)$ 
ranging between approximately $(0.3056,0.7459)$ 
(at $S_{\rm L}\simeq 0.5054$) and $(1/3,0)$ (at $S_{\rm L}=8/9$).  
As noted previously, beyond $S_{\rm L}=8/9$ there are no entangled states.  
As the inset of Fig.~\ref{Fig:ARSl} shows, the parameter $a$ can be 
regarded as a continous function of parameter 
$\mygamma\in[0,1]$.  The two branches of the solution, (i) and (ii), 
cross at $(S^*_{\rm L},E^*_{\rm R})\simeq(0.5054,0.3422)$; 
at this point, the states on the two branches coincide,
\begin{eqnarray}
\rho^*\simeq\begin{pmatrix} 0.372947 & 0 & 0 & 0.372947\cr
                       0  & 0.254106 & 0 & 0 \cr
		       0 & 0 & 0 & 0\cr
		       0.372947 & 0 & 0 & 0.372947
		       \end{pmatrix}.
\end{eqnarray}
Just as in the case of entanglement of formation vs.~linear entropy,
the density matrix is continuous at the transition between branches.

We remark that the curve generated by the states 
$\{{\rm MEMS}\!:\!E_{\rm F},S_{\rm L}\}$, 
when plotted on the $E_{\rm R}$ vs.~$S_{\rm L}$ plane, 
falls just slightly below that generated by the states 
$\{{\rm MEMS}\!:\!E_{\rm R},S_{\rm L}\}$ 
for $S_{\rm L}\agt 0.5054$ 
(and coincides for smaller values of $S_{\rm L}$). 
We also remark that the parameter $\mygamma$ turns out to be the 
concurrence $C$ of the states, so that Fig.~\ref{Fig:ARSl} can be 
interpreted a plot of the concurrence of the frontier states 
vs.~their linear entropy.  By comparing this concurrence vs.~linear 
entropy curve to that in Fig.~\ref{Fig:CSl}, we find that the former 
lies just slightly below the latter for $S_{\rm L}\agt 0.5054$ 
(and the two coincide for smaller values of $S_{\rm L}$), 
the maximal difference between the two being less than $10^{-2}$. 

\begin{figure}
\vspace{0.5cm}
\centerline{
\rotatebox{0}{
	\epsfxsize=7cm
	\epsfbox{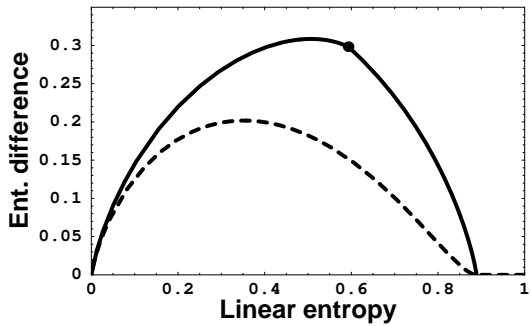}}}
\vspace{-0.2cm}
\caption{Difference in entanglement ($E_{\rm F} - E_{\rm R}$) 
 vs.~$S_{\rm L}$ for the MEMS in Eq.~(\ref{eqn:MEMS}) and Werner states.
The solid curve shows states from $\rho_{\mathrm{MEMS};E_{\rm F},S_{\rm L}}$; 
the dashed curve shows the Werner states.}
\label{Fig:dESl}
\end{figure}

It is evident that, for a given linear entropy, the relative entropies of 
entanglement for both 
$\{{\mathrm{MEMS}\!:\!E_{\rm R},S_{\rm L}}\}$ and $\{{\mathrm{MEMS}\!:\!E_{\rm F},S_{\rm L}}\}$ are significantly
less than the corresponding entanglements of formation.
In fact, for small degrees of impurity, the entanglements
of formation for the two MEMS states are quite flat; however, 
the relative entropies of entanglement fall quite rapidly.  
More specifically, for a change in linear entropy of $\Delta S_{\rm L}=0.1$ 
near $S_{\rm L}=0$ we have that 
$\Delta E_{\rm F}\approx 0.05$ (see Fig.~\ref{Fig:CSl}) and 
$\Delta E_R\approx 0.2$ (see Fig.~\ref{Fig:RSl}).  
As the curves of the states 
$\{{\rm MEMS}\!:\!E_{\rm F},S_{\rm L}\}$ and 
$\{{\rm MEMS}\!:\!E_{\rm R},S_{\rm L}\}$ are 
very close on the two planes, $E_{\rm F}$ vs.~$S_{\rm L}$ and 
$E_{\rm R}$ vs.~$S_{\rm L}$, we show, in Fig.~\ref{Fig:dESl}, 
the entanglement difference $E_{\rm F}-E_{\rm R}$ for the states 
$\{{\rm MEMS}\!:\!E_{\rm F},S_{\rm L}\}$, and compare it with the 
corresponding difference for the Werner states.  While it is clear that
$E_{\rm R} (\rho) \leq E_{\rm F} (\rho)$, for certain values of the linear 
entropy the difference turns out to be quite large, this difference being 
uniformly larger for $\{{\rm MEMS}\!:\!E_{\rm F},S_{\rm L}\}$ than for the 
Werner state; see Fig.~\ref{Fig:dESl}. This raises several
interesting questions, e.g., whether, for a given value of $E_{\rm F}$, the undistillable entanglement (the difference between the entanglement of 
formation and the relative entropy of entanglement is the lower bound)
increases as states become more mixed.

As we have seen, Werner states are not frontier states either in the case of entanglement of formation or in the case of relative entropy of entanglement. 
By contrast, as we shall see in the next section, if we measure entanglement 
via negativity, then for a given amount of linear entropy, the Werner states (as well as another Rank-3 class of states) achieve the largest 
value of entanglement. Said equivalently, the Werner states belong to $\{{\rm MEMS\!:\!N,S_{\rm L}}\}$.

\subsubsection{Negativity}
\label{sec:NSl}
\begin{figure}
\vspace{0.5cm}
\centerline{
\rotatebox{0}{
	\epsfxsize=8cm
	\epsfbox{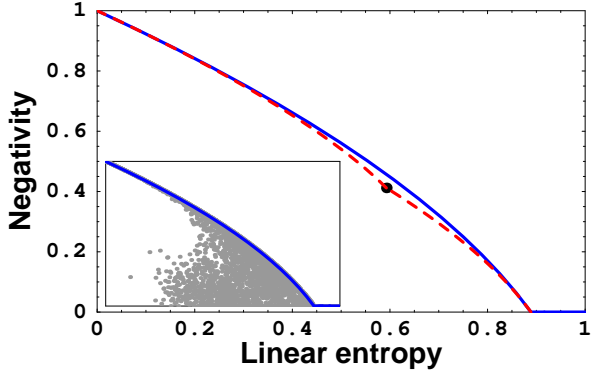}}}
\vspace{-0.2cm}
\caption{Entanglement frontier: negativity vs.~linear entropy. States on the boundary (full line) are 
$\rho^{(1)}_{{\rm MEMS}:N,S_{\rm L}}$ and 
$\rho^{(2)}_{{\rm MEMS}:N,S_{\rm L}}$. 
The dashed line comprises 
$\rho_{{\rm MEMS}:E_{\rm F},S_{\rm L}}$.
Inset: the randomly generated states confirm the location of the frontier.}
\label{Fig:NSl}
\end{figure}

In order to derive the form of the MEMS in the case of negativity, 
we again consider the density matrix of the form~(\ref{eqn:Rhomems}), 
for which it is straightforward to show that the negativity $N$ is 
given by 
\begin{eqnarray}
N=\max\{0, \sqrt{(a-b)^2+\mygamma^2}-(a+b)\}.
\label{eqn:Negativity}
\end{eqnarray}
Furthermore, because we aim to find the entanglement frontier, 
we can simply restrict our attention to states satisfying 
$N>0$, i.e., to states that are entangled~\cite{Peres96_Horodecki396}.
Then, by making $N$ stationary at fixed $S_{\rm L}$ and with the constraint
$2x+a+b+\mygamma=1$, 
we find two one-parameter families of stationary states (in addition to the Rank-2 MEMS, which are common to all three entanglement measures).
The parameters of the first family obey
\begin{eqnarray}
a=b=x,\ \mygamma=1-4x.
\end{eqnarray}
When expressed in terms of parameter $\mygamma$, the density matrix takes the form
\begin{eqnarray}
\rho^{(1)}_{\mbox{\tiny MEMS:$N\!,\!S_{\rm L}$}}
=       \begin{pmatrix}
  \frac{1+\mygamma}{4} & 0 & 0 & \frac{\mygamma}{2} \cr
               0  & \frac{1-\mygamma}{4} & 0 & 0 \cr
               0  & 0 &  \frac{1-\mygamma}{4} & 0 \cr
               \frac{\mygamma}{2} & 0 & 0 & \frac{1+\mygamma}{4}
	       \end{pmatrix}, 
\label{eqn:MEMS_N_SL1}
\end{eqnarray}
which are precisely the Werner states in Eq.~(\ref{eqn:Werner}).
For the second solution, the parameters obey
\begin{equation}
a\!=\!\frac{4\!-\!2\sqrt{3\mygamma^2\!+\!1}}{6}, b\!=\!0, 
x\!=\!\frac{1\!+\!\sqrt{3\mygamma^2\!+\!1}}{6}\!-\!
\frac{\mygamma}{2}.
\end{equation}
When expressed in terms of parameter $\mygamma$, the density matrix takes the form
\begin{equation}
\label{eqn:MEMS_N_SL2}
\rho^{(2)}_{\mbox{\tiny MEMS:$N\!,\!S_{\rm L}$}}
\!=\!
   \begin{pmatrix}
    \frac{1+\sqrt{3\mygamma^2\!+\!1}}{6} \!\!&\!\! 0 \!&\! 0\! \!& \frac{\mygamma}{2} \cr
               0 \!\!&\!\! \frac{4-2\sqrt{3\mygamma^2\!+\!1}}{6} \!&\! 0 \!\!& 0 \cr
               0 \!\!&\!\! 0 \!& \! 0\! \!& 0 \cr
               \frac{\mygamma}{2} \!\!&\!\! 0 \!&\! 0\! \!& \frac{1+\sqrt{3\mygamma^2\!+\!1}}{6}
	       \end{pmatrix}\!.
	       \end{equation} 
We remark that the two solutions give the same bound on the negativity for
a given value of linear entropy.
The resulting frontier in the negativity vs.~linear entropy plane is shown in Fig.~\ref{Fig:NSl}. 
We have shown by numerical exploration~\cite{FOOT:NSL} that the family 
states found in the previous paragraph do indeed achieve the maximal 
negativity. The results of this exploration are indicated by points in 
the inset in Fig.~\ref{Fig:NSl}, and compelling evidence for the 
maximality of the negativity is furnished by the fact that no points 
lie beyond the frontier.

Thus, the states
$\{{\rm MEMS}\!:\!N,S_{\rm L}\}$ 
on the boundary include, up to local unitary transformations,
both Werner states in Eq.~(\ref{eqn:MEMS_N_SL1}) and states in Eq.~(\ref{eqn:MEMS_N_SL2}).  We also plot in Fig.~\ref{Fig:NSl} 
the curve belonging to 
$\{{\rm MEMS}\!:\!E_{\rm F},S_{\rm L}\}$; 
note that it falls slightly below the curve associated with 
$\{{\rm MEMS}\!:\!N,S_{\rm L}\}$ 
and that it has a cusp, due to 
the structure of the states, at the value $2/3$ for the parameter
$\mygamma$ in Eq.~(\ref{eqn:MEMS}).  Here, we see that maximally 
entangled mixed states change their form when we adopt a different 
entanglement measure.

\subsection{Entanglement-versus-von-Neumann-entropy frontiers}
We continue this section by choosing to measure mixedness in terms of 
the von Neumann entropy, and comparing the frontier states for various 
measures of entanglement.

\subsubsection{Entanglement of formation}
\label{sec:EfSv}
To find this frontier, we consider states of the form~(\ref{eqn:Rhomems}), 
and compute for them the concurrence and the von Neumann entropy: 
\begin{subequations}
\begin{eqnarray}
\label{eqn:C}
\!\!\!\!\!\!\!\!\!\!\!\!\!C\!\!\!
&=&
\!\!\!\mygamma-2\sqrt{ab}, \\
\label{eqn:S_V}
\!\!\!\!\!\!\!\!\!\!\!\!\!S_{\rm V}\!\!\!
&=&
\!\!\!-a \log a \! -\! b \log b \!-\! x \log x \!-\!(x\!\!+\!\!\mygamma) \log(x\!\!+\!\!\mygamma).
\end{eqnarray}%
\label{eqn:CnSv}%
\end{subequations}%
Note that the parameters obey the normalization constraint 
$2x\!+\!a\!+\!b\!+\!\mygamma=1$.

As we remarked previously, the Rank~2 MEMS is always a candidate.  
For the Rank-3 case, we can set $b=0$ in Eq.~(\ref{eqn:CnSv}).  
By maximizing $C$
at fixed $S_{\rm V}$, we find a stationary solution: \\
(i)~$\mygamma=C$, $x=(4\!-\!3C\!-\!\sqrt{4\!-\!3C^2})/6$, and $a=(\sqrt{4\!-\!3C^2}\!-\!1)/3$; the resulting density matrix is
\begin{eqnarray}
\label{eqn:rhoI}
\rho_{\rm i}=\begin{pmatrix}
              \frac{4-\sqrt{4-3 C^2}}{6} \!&\! 0 \!&\! 0 & \frac{C}{2} \cr
              0  \!& \!\frac{\sqrt{4-3C^2}-1}{3} \!&\! 0 & 0 \cr
              0  \!& \!0\! &\! 0 & 0 \cr
              \frac{C}{2} \!& \!0 \!&\! 0 & \frac{4-\sqrt{4-3C^2}}{6} \cr
	      \end{pmatrix}\!. 
\end{eqnarray}%%
For the Rank-4 case ($b\ne0$), the stationarity condition can be shown to
be 
\begin{subequations}
\begin{eqnarray}
u\log(u)& =& w\log(w), \\
 2u\log(u)&=&(u+w)\log(v),
\end{eqnarray}
\end{subequations}%%
where 
$u\equiv\sqrt{a/(x\!+\!\mygamma)}$, 
$v\equiv\sqrt{x/(x\!+\!\mygamma)}$, and
$w\equiv\sqrt{b/(x\!+\!\mygamma)}$.
There are two solutions, due to the two-to-one property of the 
function $z\log z$ for $z\in(0,1)$. The first one is ($u=v=w$). \\
(ii)~$a=b=x=(1\!-\!C)/6$, and $\mygamma=(1\!+\!2C)/3$, which
can readily be seen to be a Werner state as in Eq.~(\ref{eqn:Werner}) 
or, equivalently, 
\begin{eqnarray}
\label{eqn:rhoII}
\rho_{\rm ii}= \begin{pmatrix}
              \frac{2+C}{6} & 0 & 0 & \frac{1+2C}{6} \cr
              0  & \frac{1-C}{6} & 0 & 0 \cr
              0  & 0 & \frac{1-C}{6} & 0 \cr
              \frac{1+2C}{6} & 0 & 0 & \frac{2+C}{6} \cr
	      \end{pmatrix}. 
\end{eqnarray}%%
Being the concurrence, $C$ is restricted to the interval $[0,1]$.
The second solution is transcendental, but can be solved numerically. 
 
\begin{figure}
\vspace{0.5cm}
\centerline{
\rotatebox{0}{
        \epsfxsize=9cm
        \epsfbox{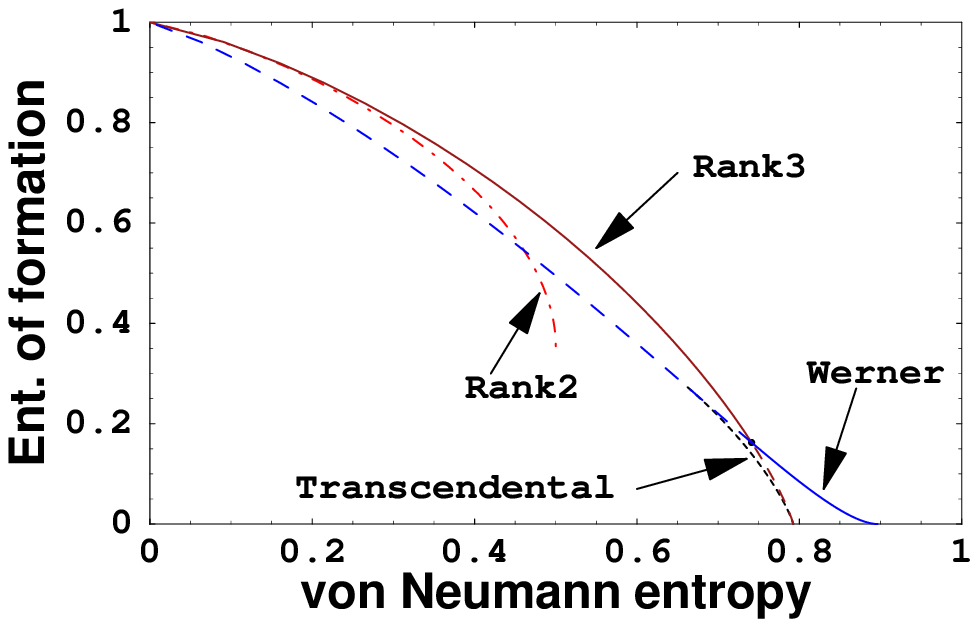}
}
}
\vspace{-0.1cm}
\centerline{
\rotatebox{0}{
        \epsfxsize=9cm
        \epsfbox{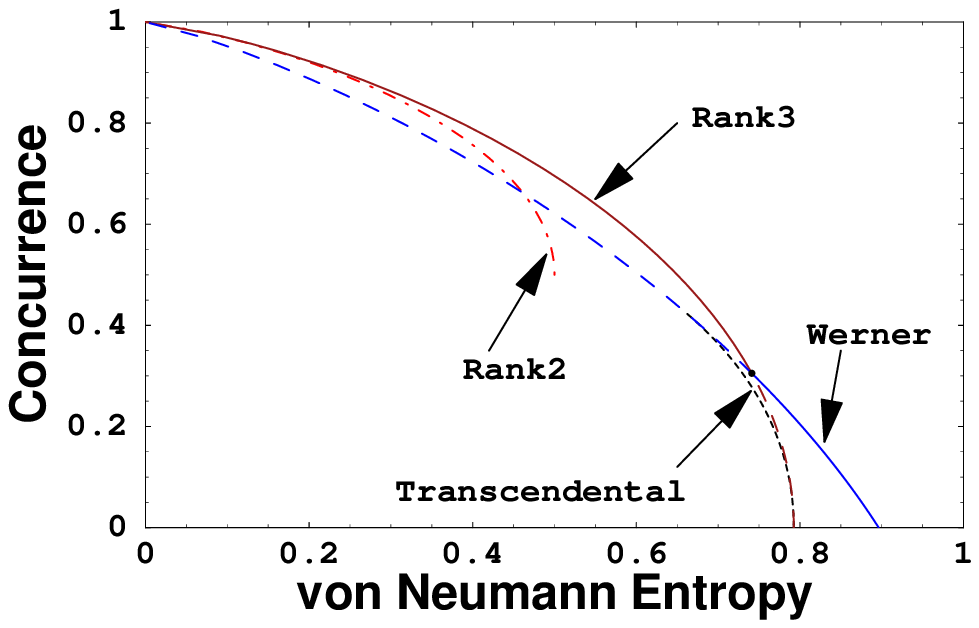}
}
}
%\centerline{$S$}
\vspace{-0.2cm}
\caption{Entanglement frontiers. Upper panel: entanglement of formation vs.~von Neumann entropy. Lower panel: concurrence vs.~von Neumann entropy. The branch structure is described in the text.
}
\label{Fig:CSv}
\end{figure}

In Fig.~\ref{Fig:CSv} we compare the four possible candidate solutions, 
and find that the global maximum is composed of only (i) and~(ii).
We summarize the states at the frontier as follows:
\begin{eqnarray}
\rho_{\mathrm{MEMS}:E_{\rm F},S_{\rm V}}=     
\begin{cases}
	\rho_{\rm ii}, 
	    &{\rm for}\ 0\le C\le C^*;\cr
	\rho_{\rm i},
	    &{\rm for}\ C^*\le C\le 1.\cr
	    \end{cases}
\label{eqn:MEMS_C_Sv}
\end{eqnarray}
Note the crossing point at $\big(C,S_{\rm V}\big)=\big( C^*,S_{\rm V}(C^*)\big)$, at which extremality is exchanged, so the true frontier consists of two branches. It is readily seen that $C^*$ is the solution of the equation $S_{\rm V}(\rho_{\rm i}(C))=S_{\rm V}(\rho_{\rm ii}(C))$,
and the approximate numerical values of $C^*$ and the corresponding $S_{\rm V}^*$ are 
$0.305$ and $0.741$, respectively.

The resulting form of MEMS states is peculiar, in that, even at the crossing point of two branches on the entanglement-mixedness plane, the forms of matrices
on the two branches are not equivalent 
(one is Rank~3, the other Rank~4).  
This is in contrast to the 
$\{{\rm MEMS}\!:\!E_{\rm F},S_{\rm L}\}$.  
This peculiarity can be partially understood from the plot
of the two mixedness measures, Fig.~\ref{Fig:SvSl}: as the value of the von Neumann entropy rises, there are fewer and fewer Rank-3 entangled states,
and above some threshold, no more Rank-3 states exist, let alone entangled Rank-3 states. There are, however, still entangled states of Rank 4. Hence, if Rank-3 states attain higher entanglement than Rank-4 states do when the entropy is low, a transition must occur between MEMS states of Rank~3 and Rank~4. 

From Fig.~\ref{Fig:CSv} it is evident that beyond a certain value of the von Neumann entropy no entangled states exist.  This value can be readily obtained
by considering the MEMS state~(\ref{eqn:rhoII}) at $C=0$,
\begin{eqnarray}
 \begin{pmatrix}
              \frac{1}{3} & 0 & 0 & \frac{1}{6} \cr
              0  & \frac{1}{6} & 0 & 0 \cr
              0  & 0 & \frac{1}{6} & 0 \cr
              \frac{1}{6} & 0 & 0 & \frac{1}{3} \cr
	      \end{pmatrix}, 
\end{eqnarray}%%
for which $S_{\rm V}= -(1/2)\log_4(1/12)\approx 0.896$.

As an aside, we mention a tantalizing but not yet fully developed analogy 
with thermodynamics~\cite{Horodecki398}. In this analogy, one associates 
entanglement with energy and von Neumann entropy with entropy, and it is 
therefore tempting to regard the MEMS just derived as the analog of 
thermodynamic equilibrium states. If we apply the Jaynes principle to an 
ensemble in equilibrium with a given amount of entanglement then the most 
probable states are those MEMS shown above.

\subsubsection{Relative entropy of entanglement}
\label{sec:ReSv}
\begin{figure}
\vspace{0.5cm}
\centerline{
\rotatebox{0}{
	\epsfxsize=7cm
	\epsfbox{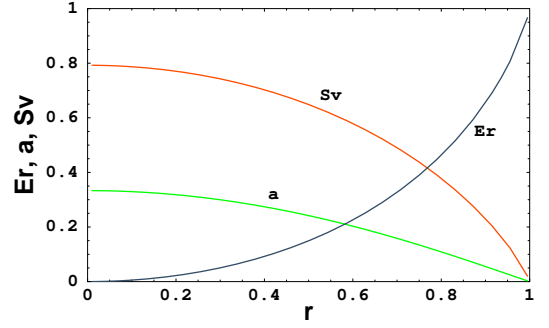}}}
\vspace{-0.2cm}
\caption{Dependence of $E_{\rm R}$, $S_{\rm V}$ and $a$ on $\mygamma$ for the Rank-3 maximal states.}
\label{Fig:ASvErOfR}
\end{figure}

Let us now find the frontier states for the case of 
relative entropy of entanglement. To do this, we first consider the 
Rank-3 states in Eq.~(\ref{eqn:rank2n3}), for which the relative entropy 
is given by Eq.~(\ref{eqn:ER3}). For these states, the von Neumann entropy is 
given by
\begin{equation}
S_{\rm V}\!=\!-\frac{1\!\!-\!\!a\!\!+\!\!\mygamma}{2}\!\log\!\!\frac{1\!\!-\!\!a\!\!+\!\!\mygamma}{2}-a\log a-\frac{1\!\!-\!\!a\!\!-\!\!\mygamma}{2}\!\log \!\!\frac{1\!\!-\!\!a\!\!-\!\!\mygamma}{2},
\label{eqn:SV3}
\end{equation}
where the $\log$ function is taken to be base four. Even though the
$\log$ functions in $S_{\rm V}$ and $E_{\rm R}$ use different
bases, the stationary condition for the parameters $\mygamma$ and $a$ does
not change, because the difference can be absorbed by a rescaling of the constraint-enforcing Lagrange multiplier. Thus, in maximizing $E_{\rm R}$ at 
fixed $S_{\rm V}$, we arrive at the stationarity condition
\begin{equation}
\label{eqn:stationaryReSv}
\log\!\frac{(\!1\!\!+\!\!a\!)^2\!
-\!\mygamma^2}{(\!1\!\!+\!\!a\!)^2}
\log\!\frac{1\!\!-\!\!a\!\!-\!\!\mygamma}{1\!\!-\!\!a\!\!
+\!\!\mygamma}\!=\!\log\!\frac{1\!\!
+\!\!a\!\!+\!\!\mygamma}{1\!\!
+\!\!a\!\!-\!\!\mygamma}\log\!\frac{(\!1\!\!
-\!\!a\!)^2\!-\!\mygamma^2}{4a^2}.
\end{equation} 
We can solve for the parameter $a$ as a function of $r\in[0,1]$, 
at least numerically; the result is shown in Fig.~\ref{Fig:ASvErOfR}, 
along with $S_{\rm V}$ and $E_{\rm R}$.
\begin{figure}
\vspace{0.5cm}
\centerline{
\rotatebox{0}{
	\epsfxsize=8cm
	\epsfbox{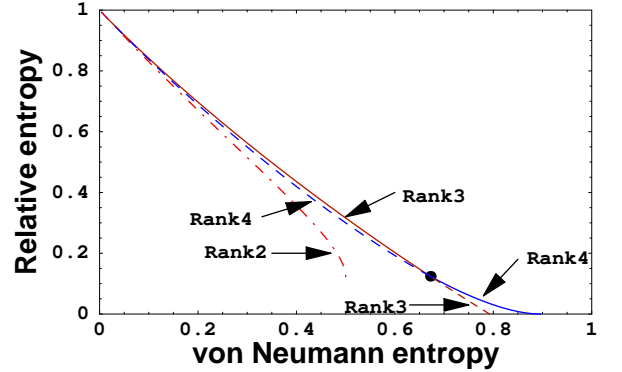}}}
\vspace{-0.2cm}
\caption{Entanglement frontier: 
relative entropy of entanglement vs.~von Neumann entropy. 
The solid curve is the frontier. 
The branch structure is described in the text.}
\label{Fig:ReSV}
\end{figure}

Turning to the Rank-4 case, it is straightforward, if tedious, to show 
that the Werner states satisfy the corresponding stationarity conditions. 
In order to ascertain which rank gives the MEMS for a given $S_{\rm V}$, 
we compare the stationary states of Rank 2, 3, and 4 in Fig.~\ref{Fig:ReSV}. 
Thus, we see that for $S_{\rm V}\le S^*_{\rm V}\simeq 0.672$ the frontier 
states are given by the Rank-3 states, whereas for $S_{\rm V}\ge S_{\rm V}^*$ 
the frontier states are given by the Werner states~(\ref{eqn:Werner})  
with the parameter $\mygamma$ ranging from approximately 0.6059 down to 0.  
At the crossing point, $(S^*_{\rm V},E_{\rm R}^*)\simeq(0.672,0.124)$, the 
MEMS undergo a discontinuous transition; recall that we encountered a 
similar phenomenon in the case of entanglement of formation vs.~von Neumann 
entropy. 

\subsubsection{Negativity}
We saw in Sec.~\ref{sec:NSl} that there is a pair of families of MEMS which differ in rank but give the identical frontier in the $N$ vs.~$S_{\rm L}$ plane. %%%
% In Secs.~\ref{sec:EfSv} and \ref{sec:ReSv} we also saw that there exists a
% discontinuous transition of MEMS states when we use von Neumann entropy as the
% mixedness measure. 
%%%
It is interesting to see what happens for the combination of negativity and von Neumann entropy. 

Once again, we begin with states of form~(\ref{eqn:Rhomems}), for which the negativity
and the von Neumann entropy are given in Eqs.~(\ref{eqn:Negativity}) and~(\ref{eqn:S_V}), respectively. By making $N$ stationary at fixed $S_{\rm V}$, we are able to find only one solution (in addition to the Rank-2 candidate): $a=b=x$. Expressing the resulting density matrix, as we may, in terms of the single parameter $\mygamma$, we arrive at the following candidate for the frontier states:
\begin{eqnarray}
\rho_{{\rm MEMS}:N,S_{\rm V}}
=       \begin{pmatrix}\frac{1+\mygamma}{4} & 0 & 0 & \frac{\mygamma}{2} \cr
               0  & \frac{1-\mygamma}{4} & 0 & 0 \cr
               0  & 0 &  \frac{1-\mygamma}{4} & 0 \cr
               \frac{\mygamma}{2} & 0 & 0 & \frac{1+\mygamma}{4}
	       \end{pmatrix}, 
\label{eqn:MEMS_N_SV}
\end{eqnarray}
where $0\le\mygamma\le 1$, i.e., the Werner states.   

\begin{figure}
\vspace{0.5cm}
\centerline{
\rotatebox{0}{
        \epsfxsize=7cm
        \epsfbox{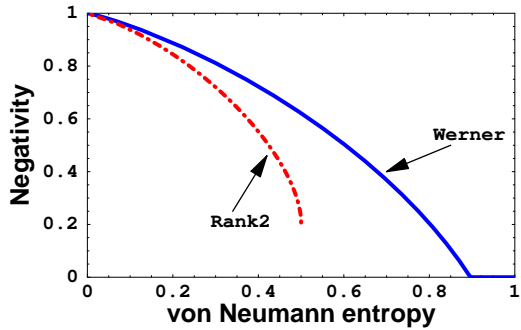}
}
}
%\centerline{$S$}
\vspace{-0.2cm}
\caption{Entanglement frontier: 
negativity vs.~von Neumann entropy.  
The solid curve is the frontier.  
The broken curve represents the Rank-2 candidate states.
}
\label{Fig:NSv}
\end{figure}
The resulting frontier in the negativity vs.~von-Neumann-entropy 
plane is shown in Fig.~\ref{Fig:NSv} which, for comparison, also 
shows the curve for the Rank-2 candidate. 
\section{Concluding Remarks}
\label{SEC:Conclude}
\begin{figure}
\vspace{0.5cm}
\centerline{
\rotatebox{0}{
        \epsfxsize=8cm
        \epsfbox{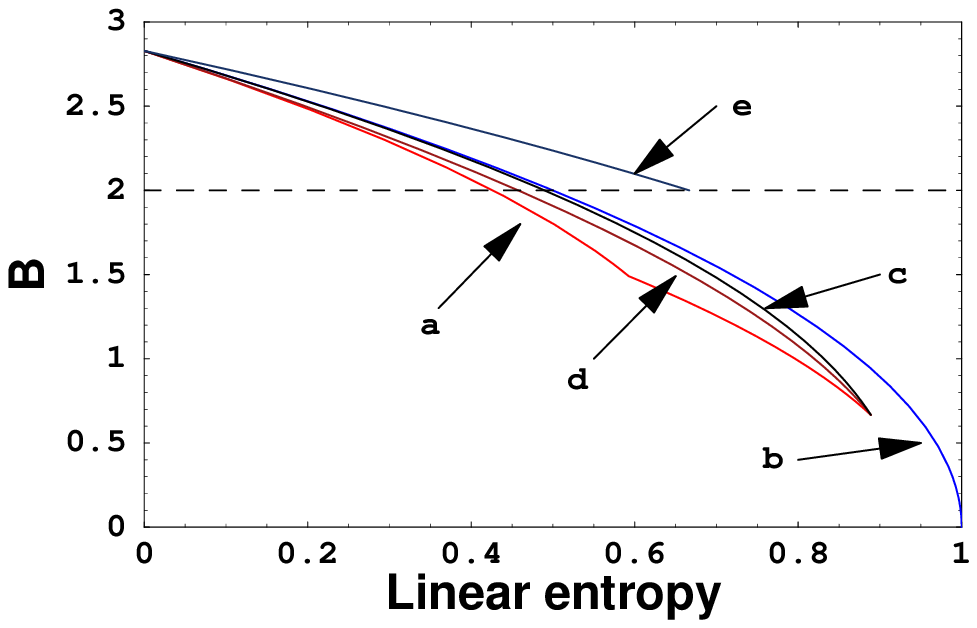}
}
}
%\vspace{-0.2cm}
\centerline{
\rotatebox{0}{
        \epsfxsize=8cm
        \epsfbox{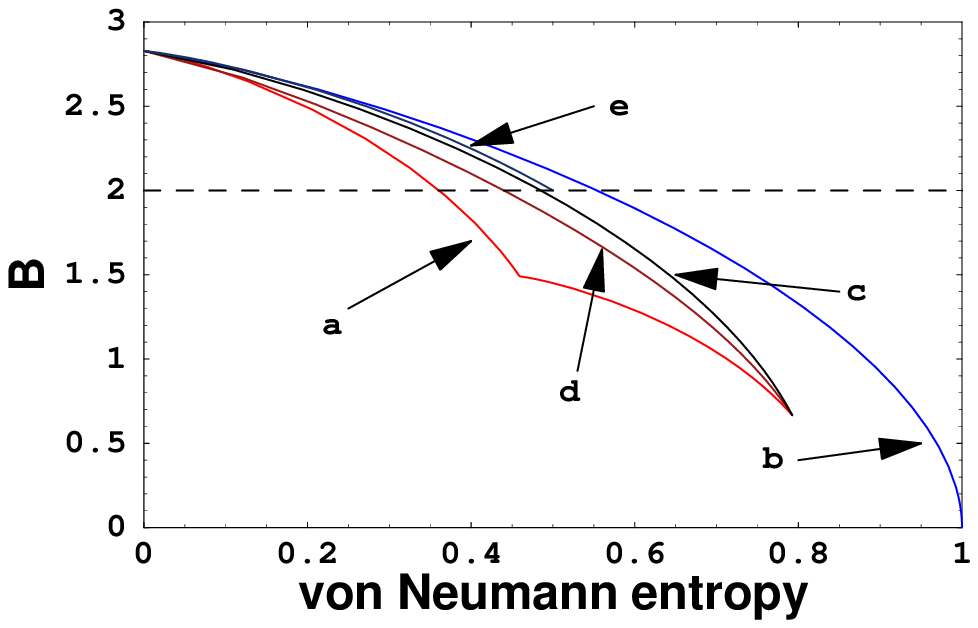}
}
}
%\centerline{$S$}
\vspace{-0.2cm}
\caption{Violation of Bell's inequality 
for various families of states: 
(a)~$\{{\rm MEMS}\!:\!E_{\rm F},S_{\rm L}\}$, 
(b)~Werner states, 
(c)~$\rho^{(2)}_{\mbox{\tiny MEMS:$N\!,\!S_{\rm L}$}}$, 
(d)~$\rho_{\rm i}$ in Eq.~(\ref{eqn:rhoI}), and 
(e)~the Rank-2 Bell diagonal states in Eq.~(\ref{eqn:Rank2Bell}).
}
\label{Fig:Bell}
\end{figure}
In this Paper we have determined families of maximally entangled 
mixed states (MEMS, i.e., frontier states, which possess the maximum 
amount of entanglement for a given degree of mixedness). These states 
may be useful in quantum information processing in the presence of 
noise, as they have the maximum amount of entanglement possible for a 
given mixedness. We considered various measures of entanglement 
(entanglement of formation, relative entropy, and negativity) and 
mixedness (linear entropy and von Neumann entropy). 

We found that the form of the MEMS depends heavily on the measures used. 
Certain classes of frontier states (such as those arising with either 
entanglement of formation or relative entropy of entanglement vs.~the 
von Neumann entropy) behave discontinously at a specific point on the entanglement-mixedness frontier. Under most of the settings considered, we 
have been able to explicitly derive analytical forms for the frontier states.

For entanglement of formation and relative entropyand for most values of 
mixedness, we have found that the Rank-2 and Rank-3 MEMS have more 
entanglement than Werner states do. On the other hand, at fixed entropy 
no states have higher negativity than Werner states do. At small amounts 
of mixedness, the 
$\{{\rm MEMS}\!:\!E_{\rm F},S_{\rm L}\}$ states 
\lq\lq lose\rq\rq\ entanglement with increasing mixedness at a substantially 
lower rate than do the Werner states. However, when the entanglement is 
measured by the relative entropy, the difference in loss-rate is 
significantly smaller. 

From Eq.~(\ref{eqn:LSDecomp}) it is tempting to assert that in the case 
of LS entanglement vs.~mixedness the frontier states should be the same 
as those in the case of entanglement of formation vs.~mixedness. However, 
as we do not know whether Eq.~(\ref{eqn:Ansatz}) (up to local unitary 
transformations) exhausts all maximal states for LS entanglement,  
further investigation of this point is needed. 

Having characterized the MEMS for various measures, it is worthwhile 
considering them from the perspective of Bell-inequality violations.  
To quantify the violation of Bell's inequality, it is useful to 
consider the quantity 
\begin{equation}
B\equiv\max_{\vec{a},\vec{a}',\vec{b},\vec{b}'}
\big\{E(\vec{a},\vec{b})+E(\vec{a},\vec{b}')
+E(\vec{a}',\vec{b})-E(\vec{a}',\vec{b}')\big\},
\end{equation} 
where $E(\vec{a},\vec{b})\equiv \langle\vec{\sigma}\cdot \vec{a}\otimes\vec{\sigma}\cdot\vec{b}\rangle$ and the vectors $\vec{a}$ 
and $\vec{a}'$ ($\vec{b}$ and $\vec{b}'$) are two different measuring 
apparatus settings for observer A (observer B). If $B>2$ then the 
corresponding state violates Bell's inequality.
For the density matrix of the form~(\ref{eqn:Rhomems}) 
it is straightforward~\cite{Aravind95} to show that the quantity
$B$ is given by
\begin{eqnarray}
B=2\sqrt{\Big(4(x+\frac{\mygamma}{2})-1\Big)^2+\mygamma^2}.
\end{eqnarray}
In Fig.~\ref{Fig:Bell}, we plot $B$ vs.~linear and von Neumann entropies for several families of frontier states. 
As a comparison, we also draw the violation by the 
following Rank-2 state (which is diagonal in the Bell basis):
\begin{eqnarray}
\label{eqn:Rank2Bell}
\rho_{\rm B}\equiv\mygamma|\phi^+\rangle\langle\phi^+|+(1-\mygamma)|\phi^-\rangle
\langle\phi^-|, \ \mygamma\in[0,1].
\end{eqnarray}
This state, although not belonging to any of the families of frontier 
states derived previously, turns out to achieve the maximum possible 
violation, as a function of linear entropy. On the other hand, the 
Werner states appear to achieve maximal violation in the case of von 
Neumann entropy.  Ekert's application of Bell's inequalities to quantum 
cryptography~\cite{Ekert91}, together with the discussions of the present 
paragraph, suggests that MEMS may be relevant to quantum communication.

Another natural application for which entanglement is known to be a 
critical resource is quantum teleportation. How do these frontier MEMS 
teleport, compared with the Werner and Rank-2 Bell diagonal states?  If we 
restrict our attention to high purity situations (i.e., to states with only 
a small amount of mixedness) then it is straightforward to show that, e.g., 
$\{{\rm MEMS}\!:\!E_{\rm F},S_{\rm L}\}$ 
states teleport average 
states better than the Werner states do, but worse than the Rank-2 Bell 
diagonal state does. Part of the explanation for this behavior is that 
standard teleportation is optimized for using Bell states as its core resource. 

It is also interesting to note that for certain combinations of entanglement and mixedness measures, as well as the Bell inequality violation, the Rank-2 candidates fail to furnish MEMS. Thus, these states seem to be less useful than other MEMS. However, from the perspective of distillation, these states are exactly quasi-distillable~\cite{Horodecki399,VerstraeteDehaeneDeMoor01}, and can be useful in the presence of noise because they can be easily distilled into Bell states. 

\section*{Acknowledgements}
This material is based on work supported by NSF EIA01-21568 (TCW, PMG and PGK) 
and by the U.S.~Department of Energy, Division of Material Sciences, under 
Award No.~DEFG02-91ER45439, through the Frederick Seitz Materials Research 
Laboratory at the University of Illinois at Urbana-Champaign (TCW and PMG).  
KN and WJM acknowledge financial support from the European projects QUIPROCONE 
and EQUIP.  PMG gratefully acknowledges the hospitality of the Unversity of 
Colorado at Boulder, where part of this work was carried out.  TCW gratefully acknowledges the receipt of a Mavis Memorial Fund Scholarship. 

\appendix

\section{Number of negative eigenvalues of the partial transpose of $\rho$}
\label{app:OneNeg}
In this appendix we address the result that for $C^2\otimes C^2$ 
systems the partial transpose of any density matrix $\rho$ has at 
most one negative eigenvalue. In fact, we shall consider the result 
from two perspectives.

First, we build upon results (Theorem 3, in particular) contained in Ref.~\cite{VerstraeteDehaeneDeMoor01},  from which it follows that
it is sufficient to consider 
(i)~Bell diagonal states and 
(ii)~states of the form 
\begin{equation}
\begin{pmatrix}
a+c & 0 & 0 & d \cr
0  & 0 & 0 & 0\cr
0  & 0 & b-c & 0\cr
d & 0 & 0 &a-b\end{pmatrix}.
\end{equation}
For the latter case, straightforward calculation shows that the 
partially transposed matrix can have one negative eigenvalue when 
$d\ne 0$ and that it does not have negative eigenvalues when $d=0$.  
For the former case, suppose that the Bell diagonal state has the 
four eigenvalues
$\lambda_1$, $\lambda_2$, $\lambda_3$, $\lambda_4$, 
in nonincreasing order.  Then it is straightforward to see that the 
corresponding partially transposed matrix has the four eigenvalues 
$( \lambda_1+\lambda_2+\lambda_3-\lambda_4)/2$, 
$( \lambda_1+\lambda_2-\lambda_3+\lambda_4)/2$, 
$( \lambda_1-\lambda_2+\lambda_3+\lambda_4)/2$,
$(-\lambda_1+\lambda_2+\lambda_3+\lambda_4)/2$.  
Thus, it can have at most one negative eigenvalue. 

A second perspective is provided by first invoking the LS decomposition~(\ref{eqn:LSdec}) and then making a Schmidt 
decomposition, via the local unitary transformation 
$U_A\otimes U_B$, of the pure entangled part $|\psi_e\rangle$.  
In this way, the pure part becomes 
\begin{eqnarray}
U_A\otimes U_B |\psi_e\rangle=\sqrt{p}|00\rangle+\sqrt{1-p}|11\rangle,
\end{eqnarray}
where $1/2\le p\le 1$. Meanwhile, $\rho$ is transformed into
\begin{eqnarray}
\rho'&=& 
(U_A\otimes U_B) 
\ \rho \ (U_A^\dag\otimes U_B^\dag) 
\nonumber \\
&= &\lambda\rho_s'\! +\!
      (1\!-\!\lambda)\!\!\begin{pmatrix}
                    \!
                  p \!&\! 0 \!&\! 0 \!&\!\sqrt{p(1\!-\!p)}\! \cr
                  \!0 \!&\! 0 \!&\! 0 \!&\! 0 \!\cr
                  \!0 \!&\! 0 \!&\! 0 \!&\! 0 \!\cr
                 \!\sqrt{p(1\!-\!p)} \!&\! 0 \!&\! 0\! &\! 1\!-\!p \!\cr
		 \end{pmatrix},
\end{eqnarray}
where 
$\rho_s'\equiv 
U_A\otimes U_B 
\,\rho_s \, 
U_A^\dag\otimes U_B^\dag$
is still separable, and its partial transpose remains positive semi-definite.

Taking, then, the partial transpose, the density matrix is transformed into
\begin{eqnarray}
\label{eqn:rhoTB}
{\rho'}^{{\rm T}_B} 
\!\! =\! 
\lambda{\rho_s'}^{{\rm T}_B} 
\!\!+\!  
    (1\!\!-\!\!\lambda)\! \begin{pmatrix}
                  p \!&\!\! 0 \!\!&\!\! 0 \!\!&\!\!0 \cr
                  0 \!&\!\! 0 \!\!&\!\!\sqrt{p(1\!-\!p)} \!\!&\!\! 0 \cr
                  0 \!&\!\! \sqrt{p(1\!-\!p)} \!\!&\!\! 0 \!\!&\!\! 0 \cr
                  0 \!&\!\! 0 \!\!&\!\! 0 \!\!&\! \!1\!-\!p\cr
		  \end{pmatrix}\!,
\end{eqnarray}
and we note that the last matrix has 
eigenvalues 
$\lambda_1\!=\!p$, 
$\lambda_2\!=\!\sqrt{p(1\!-\!p)}$, 
$\lambda_3\!=\!1\!-\!p$, 
$\lambda_4\!=\!-\sqrt{p(1\!-\!p)}$ 
in (as $p\ge 1/2$) nonincreasing order.  
As the Hermitian matrix ${\rho_s'}^{{\rm T}_B}$ 
retains positive-semi-definiteness, we can employ a
well-known result in matrix analysis~\cite{Horn_Johnson} that for $n\times n$ Hermitian matrices $A$ and $B$, with $B$ being positive semi-definite, 
the eigenvalues of $(A+B)$ and $A$, when arranged in non-ascending order, obey
\begin{eqnarray}
\lambda_k(A+B)\ge\lambda_k(A) 
\quad \mbox{for} \ k=1,2,\ldots,n.
\end{eqnarray}
Hence, identifying $B$ with 
$\lambda{\rho_s'}^{{\rm T}_B}$ and $A$ 
with the product of $(1\!-\!\lambda)$ 
and the matrix in Eq.~(\ref{eqn:rhoTB}), 
we immediately see that 
$  \lambda_1({\rho'}^{{\rm T}_B})
\ge\lambda_2({\rho'}^{{\rm T}_B})
\ge\lambda_3({\rho'}^{{\rm T}_B})
\ge (1-\lambda)(1-p)\ge 0$ 
and, thus, 
${\rho'}^{{\rm T}_B}$ (or equivalently 
${\rho }^{{\rm T}_B}$) 
can have at most one negative eigenvalue.  
Thus, the negativity for $C^2\otimes C^2$ 
systems can then be written as 
$N=2\max\{0,-\lambda_4(\rho^{{\rm T}_B})\}$.


\begin{thebibliography}{99}
\bibitem{FOOT:entropy}
By entropy we mean a measure of how mixed a state is 
(i.e., its mixedness).  We shall focus on two measures: 
the linear entropy and the von Neumann entropy, 
defined in Sec.~\ref{SEC:MeasOfMix}.
\bibitem{REF:SepEnt}
A separable (or unentangled) state (bi-partite) 
$\rho^{\rm AB}_{\rm s}$ can be expressed as
$\rho_{\rm s}^{\rm AB}=
\sum_{i}p_i \ \rho^{\rm A}_i\otimes\rho^{\rm B}_i$,
whereas an entangled state has no such decomposition.
\bibitem{REF:PartialT} 
Suppose a bi-partite density matrix $\rho$ 
is expressed in the following form:
$\rho=
\sum_{\{i,j,k,l\}}
\rho_{ij;kl}\,
|e^{\rm A}_i\otimes e^{\rm B}_j\rangle
\langle e^{\rm A}_k\otimes e^{\rm B}_l|$.
Then the partial transpose $\rho^{{\rm T}_B}$ 
of the density matrix $\rho$ is defined via 
\begin{equation}
\rho^{{\rm T}_B}
\equiv
\sum_{\{i,j,k,l\}}
\rho_{il;kj}\,
|e^{\rm A}_i\otimes e^{\rm B}_j\rangle
\langle e^{\rm A}_k\otimes e^{\rm B}_l|.
\nonumber
\end{equation}
We remark that the partial transpose depends on the basis chosen  
but the eigenvalues of the partial transposed matrix do not.
\bibitem{FOOT:NSL}
By numerical exploration we mean the creation of a large set of 
random density matrices, for each of which we compute the relevant 
characteristics, in this case negativity and linear entropy.
\end{thebibliography}
\end{document}